\documentclass[12pt]{article}

\usepackage{fullpage}

\usepackage{picins}

\usepackage{amsfonts}

\usepackage{graphicx}

\usepackage{mathrsfs}

\usepackage{amsmath}

\usepackage{amssymb}

\usepackage{float}

\usepackage{color}

\usepackage{indentfirst}

\newcommand{\bm}{\boldsymbol}

\title{Simultaneous Monitoring of a Large Number of Heterogeneous Categorical
Data Streams}
\author{Kaizong Bai$^1$ and
Jian Li$^1$\footnote{{Corresponding author. Email: jianli@xjtu.edu.cn}}
\\
$^1${\em\small School of Management, Xi'an Jiaotong University, Xi'an, Shaanxi, China}
}

\date{}

\begin{document}

\maketitle

\begin{abstract} \baselineskip 18pt

This article proposes a powerful scheme to monitor a large number of categorical
data streams with heterogeneous parameters or nature. The data streams considered
may be either nominal with a number of attribute levels or ordinal with some natural
order among their attribute levels, such as good, marginal, and bad. For an ordinal
data stream, it is assumed that there is a corresponding latent continuous data
stream determining it. Furthermore, different data streams may have different number
of attribute levels and different values of level probabilities. Due to high
dimensionality, traditional multivariate categorical control charts cannot be
applied. Here we integrate the local exponentially weighted likelihood ratio test
statistics from each single stream, regardless of nominal or ordinal, into a
powerful goodness-of-fit test by some normalization procedure. A global monitoring
statistic is proposed ultimately. Simulation results have demonstrated the
robustness and efficiency of our method.

\vspace{0.2cm} \noindent{{\bf Keywords}}: EWMA, Likelihood Ratio Test, Ordinal
Categorical, Statistical Process Control

\end{abstract}

\section{Introduction} \baselineskip 18pt

Statistical Process Control (SPC) has demonstrated its power in monitoring various
types of data streams. Quality control charts were firstly developed for monitoring
a single data stream, among which well-known charts include the $\bar{x}$-chart for
normally distributed data, the $p$-chart for binomial distributed data, and the
$c$-chart for Poisson counts. Then they were extended to simultaneous surveillance
for multiple data streams, known as multivariate SPC. The most famous chart for
multivariate continuous data is the Hotelling's $T^2$ chart. Recent contributions
include Zou and Qiu (2009), Wang and Jiang (2009), Capizzi and Masarotto (2011), and
so on. There are also control charts developed for multivariate categorical data.
One may refer to Patel (1973), Marcucci (1985), Lu et al. (1998), Li et al. (2012; 2014a),
and Kamranrad et al. (2017). As for monitoring multivariate Poisson counts,
recent works include Chiu and Kuo (2008), He et al. (2014) and Wang et al. (2017).
Please see Woodall
(1997), Lowry and Montgomery (1995), and Bersimis et al. (2007) for a detailed
review.

With the rapid development of information technology, together with advances in
sensory and data acquisition techniques in recent years, more and more real
applications that involve a large number of quality characteristics are to be
controlled. In these cases, traditional multivariate SPC methods, such as those
mentioned above, cannot be applied, since the number of parameters to be estimated
increases very rapidly with the dimension $p$. This is actually the curse of
dimensionality that leads to the difficulty or impossibility in parameter estimation
unless there is a very large sample. One idea can be first calculating a local test
statistic for each single data stream and then combining them in some manner to
exploit the global information over all the data streams. Along this line, recently
some efforts have been devoted to monitoring high-dimensional continuous data
streams simultaneously. Tartakovsky et al. (2006) developed a monitoring approach by
taking the maximum of the local cumulative sum (CUSUM) statistics from each single
stream. Mei (2010) proposed a monitoring scheme based on the sum of the local CUSUM
statistics from each data stream. Liu et al. (2015) extended this method and developed an
adaptive monitoring scheme by assuming that only partial data streams are observable
at one time point. In particular, Zou et al. (2015) integrated a powerful
goodness-of-fit (GOF) test with the local CUSUM statistic for each data stream.

However, all the above methods are devised for monitoring high-dimensional
continuous data streams. To the best of our knowledge, till now there is no approach
for the statistical surveillance of a large number of categorical data streams
(CDSs). Categorical data become more and more common in reality these days. In
practice, some quality characteristics are expressed as attribute levels instead of
numerical values. Moreover, due to collection cost or other difficulties, it is
difficult to obtain precise numerical values in many applications. Therefore,
monitoring schemes for high-dimensional CDSs are called for.

In fact, each CDS may be heterogenous with different parameters or nature, and the
monitoring of a large number of CDSs would bring many challenges. First, a CDS is
either nominal with some finite attribute levels only or ordinal with some natural order
among its attribute levels. Shifts in these two types of data are different. To
be specific, deviations in a nominal CDS are reflected by changes in the level
probabilities, whereas shifts in an ordinal CDS are determined by changes in its
corresponding latent continuous data stream, such as a mean shift in this latent
continuous stream. Second, different CDSs may have different number of attribute
levels and different values of level probabilities, such as binomial or multinomial,
and this would make the monitoring task more challenging. This raises two
non-trivial problems: how to devise a local test statistic with different testing
goals for nominal and ordinal CDSs, as well as how to combine these local test
statistics from heterogenous data streams to form a global monitoring statistic.

Here we take the advantage of Zou et al. (2015) to integrate the local test
statistics from each CDS into a powerful GOF test in Zhang (2002). The local
statistic from each data stream is the likelihood ratio test (LRT) statistic
equipped with the exponentially weighted moving average (EWMA) scheme, known as
exponentially weighted LRT statistic. In this article, we assume the data streams
are mutually independent, but the dependent case can also be applied, as discussed
in Zou et al. (2015). By a normalization procedure, we transform the local test
statistics from each data stream into statistics with identical scales, which
facilitates forming a global monitoring statistic with the powerful GOF test in
Zhang (2002).

The remainder of this article is organized as follow. First, the local test
statistics with different testing goals for nominal and ordinal CDSs are built in
Section 2. In Section 3, we introduce the powerful GOF test and combine the LRT with
the EWMA control scheme, and finally the global monitoring statistic for a large
number of CDSs is developed. Comparison results are shown in Section 4. A
manufacturing example is employed to demonstrate the proposed method in Section 5.
Several remarks in Section 6 conclude this article.

\section{Local Tests for Heterogenous Data Streams}


To monitor a large number of CDSs, as mentioned above, the local test statistics of
each data stream with different testing goals for nominal and ordinal CDSs should be built
first, which is the foundation of the global test statistic.

For a single nominal CDS, say the $i$th one, suppose that it takes $h_i$ possible
attribute levels. Denote the probability of an observation falling into level $j$
$(j=1,2,\dots,h_i)$ by $\pi_{ij}$, which satisfies $\sum_{j=1}^{h_i} \pi_{ij}=1$.
Given an observation $X_{it}$ at time $t$ $(t=1,2,\dots)$, the count of the
observation in level $j$, denoted by $y_{ijt}$, would be 1 or 0 and subject to
$\sum_{j=1}^{h_i}y_{ijt}=1$. For a fixed sample size $N$, the grouped count of this
data stream in level $j$ in the $k$th sample is
\begin{equation*}
n_{ijk}=\sum_{t=(k-1)N+1}^{kN}y_{ijt}.
\end{equation*}
Let ${\bf n}_{ik}=[n_{i1k},\dots,n_{ijk},\dots,n_{ih_ik}]^T$ and ${\bm
{\pi}}_i=[\pi_{i1},\dots,\pi_{ih_i}]^T$, and the Probability Mass Function(PMF) of
this nominal CDS can be expressed as
\begin{equation*}
F({\bf
n}_{ik})=\frac{N!}{\prod_{j=1}^{h_i}n_{ijk}!}\prod_{j=1}^{h_i}\pi_{ij}^{n_{ijk}}.
\end{equation*}

The monitoring of this nominal CDS is to test whether there is any change in the
probability vector ${\bm {\pi}}_i$ based on online samples, and the hypothesis can
be expressed as
\begin{equation*}
   H_0:\;{\bm {\pi}}_i= {\bm {\pi}}^{(0)}_i\ \ \mbox{versus}\ \ H_1:\; {\bm {\pi}}_i\neq {\bm
   {\pi}_i^{(0)}}.
\end{equation*}
Here  $  {\bm {\pi}}^{(0)}_i$ is the value of ${\bm {\pi}}_i$ in the null hypothesis
or the in-control (IC) state. To test the above hypothesis, the LRT  is employed.
The log-likelihood function can be written from the PMF of the multinomial
distribution based on the $k$th sample and expressed as
\begin{equation*}
    l_k({\bm {\pi}}_i)=\sum_{j=1}^{h_i} n_{ijk} \ln \pi_{ij}+\ln(N!)-\sum_{j=1}^{h_i}  \ln (n_{ijk}!).
\end{equation*}
Based on it, the maximum likelihood estimate of $\bm\pi_i$ can be calculated, and
finally the $-$2LRT statistic is
\begin{equation}
   R_{ik}=2\sum_{j=1}^{h_i}n_{ijk}\ln\frac{n_{ijk}}{N\pi_{ij}^{(0)}}.\label{f01}
\end{equation}

The difference between a nominal CDS and an ordinal one is whether there is natural
order among the attribute levels. For an ordinal CDS, say the $i$th, it is assumed
that there is a latent and unobservable continuous variable $X_i^*$, which
determines the attribute levels of $X_i$ by classifying its numerical value based on
some predefined thresholds $b_{ij}$ $(j=0, \dots, h_i)$
\begin{equation*}
    -\infty=b_{i0}<b_{i1}<\dots<b_{i,h_i-1}<b_{ih_i}=\infty.
\end{equation*}
In fact, we have
\begin{align*}
X_i\ {\mathop{\mbox{is in}}}\left\{\hspace{-0.1cm}
\begin{array}{ll}
{\mbox{level 1} }, &
{\mathrm{\ }} \quad X_i^*\in(b_{i0},b_{i1}],\\[0.1cm]
{\mbox{level 2} }, &
{\mathrm{\ }} \quad X_i^*\in(b_{i1},b_{i2}],\\[0.1cm]
{\mbox{\dots} } &
{\mathrm{\ }} \quad \dots\\[0.1cm]
{\mbox{level }h_{i} }, &
{\mathrm{\ }} \quad X_i^*\in(b_{i,h_i-1},b_{ih_i}).\\
\end{array}\right.
\end{align*}

In order to exploit the ordinal information among the attribute levels, let
$f_i(x_i^*)$ and $F_i(x_i^*)$ denote the PDF and the Cumulative Distribution
Function (CDF) of the latent continuous variable $X_i^*$, respectively. Li et al.
(2018) proposed an ordinal log-linear model
\begin{equation*}
    \ln\pi_{ij}=\beta_{i0}+\beta_{i1}\alpha_{ij}
\end{equation*}
with
\begin{equation*}
    \alpha_{ij}=\frac{1}{\pi_{ij}^{(0)}}\big[f_i\big(F_i^{-1}(c_{i,j-1}^{(0)})\big)
    -f_i\big(F_i^{-1}(c_{ij}^{(0)})\big)\big],
\end{equation*}
here $c_{ij}^{(0)}=\sum_{k=1}^j\pi_{ik}^{(0)} (j=1,2,\dots,h_i)$ and
$c_{i0}^{(0)}=0$, representing the IC cumulative probabilities up to level $j$. In
addition, $\alpha_{ij}$ can be regarded as the averaged score of $X_i^*$ in the
$j$th classification interval $(F_i^{-1}(c_{i,j-1}^{(0)}),F_i^{-1}(c_{ij}^{(0)})]$
(Li et al., 2018). If $f_i(x_i^*)$ and $F_i(x_i^*)$ are all known, $\alpha_{ij}$
could be calculated in advance. Due to the constraint $\sum_{j=1}^{h_i}\pi_{ij}=1$,
there is only one independent parameter $\beta_{i1}$ in this ordinal log-linear
model.

For this ordinal CDS, we intend to test if there is any location shift
$\delta_i\neq0$ in its latent continuous variable $X_i^*$ that brings $F_i(x_i^*)$
to $F_i(x_i^*-\delta_i)$ based on only the observed attribute levels of $X_i$. In
other words, we mean to test the hypothesis
\begin{equation*}
   H_0:\;\delta_i=0\ \ \mbox{versus}\ \ H_1:\; \delta_i\neq0.
\end{equation*}
Base on the ordinal log-linear model discussed above, it suffices to test whether
there is any shift in the coefficient $\beta_{i1}$. This is equivalent to test
whether there is a location shift $\delta_i\neq0$ in the latent CDF $F_i(x_i^*)$.
The detailed proof can be found in Li et al. (2018). As a result, the hypothesis can
be transformed to
\begin{equation*}
   H_0:\;\beta_{i1}=\beta_{i1}^{(0)}\ \ \mbox{versus}\ \ H_1:\; \beta_{i1}\neq
   \beta_{i1}^{(0)},
\end{equation*}
where $\beta_{1}^{(0)}$ is the IC value of $\beta_{1}$.

Consequently, based on the $k$th sample, the $-2$LRT statistic for testing the above
hypothesis is
\begin{equation}
    R_{ik}=\frac{(\bm
    {\alpha}_i^T\bold{n}_{ik})^2}{N\bm{\alpha}_i^T\bm{\Lambda}_i\bm{\alpha}_i},\label{f03}
\end{equation}
where $\bm\alpha_i=[\alpha_{i1},\dots,\alpha_{ih_i}]^T$ and
$\bm{\Lambda}_i=\mbox{diag}(\bm{\pi}_i^{(0)})-\bm{\pi}_i^{(0)}(\bm{\pi}_i^{(0)})^T$.

Notice that the specific forms of $\alpha_{ij}$s are required to calculate the test
statistic (\ref{f03}). However, the specific forms of the PDF  $f_i(x_i^*)$ and the
CDF $F_i(x_i^*)$ of the latent variable $X_i^*$ remain unavailable. Here we might as
well assume that by standardization the latent variable $X_i^*$ follows the standard
normal distribution with mean 0 and variance 1, $\alpha_{ij}$ would have the form
\begin{equation}
  \alpha_{ij}=\frac{1}{\pi_{ij}^{(0)}}\big[\phi\big(\Phi^{-1}(c_{i,j-1}^{(0)})\big)
  -\phi\big(\Phi^{-1}(c_{ij}^{(0)})\big)\big],\label{f03a}
\end{equation}
where $\phi$ and $\Phi$ are the PDF and CDF of the standard normal distribution,
respectively. If the latent variable is indeed normally distributed, $\alpha_{ij}$
in equation (\ref{f03a}) is definitely right. Otherwise, a better choice would be
the logistic distribution recommended by Li et al. (2014b), which has a similar
shape to the normal distribution and is robust against various types of
distributions. However, here we would not emphasize the heterogeneity introduced by
the specific forms of the latent variables of ordinal CDSs, which makes this article
more complex yet can be handled without more efforts. Instead, we assume that all
the ordinal CDSs have normally distributed latent variables.

\section{Global Monitoring Method}

Suppose that there are $p$ CDSs, including simultaneously nominal with a number of
attribute levels and ordinal with some natural order among the attribute levels. It
is assumed that all the data streams are mutually independent. Simultaneous
monitoring of a large number of CDSs is to test if there are any changes in the
probability vectors of any nominal CDSs or location shifts in the latent continuous
variables of any ordinal CDSs based on online collected samples. Without loss of
generality, it is assumed that the first $q$ CDSs are nominal and the rest $p-q$
ones are ordinal. In the out-of-control (OC) state, the probability vectors
$\bm\pi_i$ of some nominal CDSs ($i=1,\dots,q$) may deviate, and there may be also
latent location shifts $\delta_i$ in some ordinal CDSs ($i=q+1,\dots,p$). Therefore,
the monitoring problem can be formulated into testing the following hypothesis
\begin{equation}
   H_0:\; \bigcap_{i=1}^{q}( {\bm {\pi}}_i= {\bm {\pi}}_i^{(0)})\ \ \mbox{and}\bigcap_
   {i=q+1}^{p} (\delta_i=0)\ \ \ \ \mbox{versus}\ \
   H_1:\;\bigcup_{i=1}^{q}( {\bm {\pi}}_i\neq {\bm {\pi}}_i^{(0)})\ \ \mbox{or}\bigcup
   _{i=q+1}^{p}(\delta_i\neq0).\label{f04}
\end{equation}

The local test statistic of a single CDS, either nominal or ordinal, has been
developed in equations (\ref{f01}) or (\ref{f03}). Recall that the data streams
considered here are heterogenous, since nominal streams may have different number of
attribute levels and different values of level probabilities, and ordinal streams
may have different latent continuous variables. Therefore, the local test statistics
introduced above are still heterogenous. In order to overcome this difficulty, first
we equip the local test statistics with the EWMA scheme by replacing ${\bf n}_{ik}$
with its exponentially weighted version for all the CDSs
\begin{equation*}
  {\bf w}_{ik}=(1-\lambda){\bf w}_{i,k-1}+\lambda{\bf n}_{ik},
\end{equation*}
where $\lambda\in(0,1]$ is the smoothing parameter, and ${\bf w}_{i0}=N{\bm
{\pi}}_{i}^{(0)}$. This facilitates detecting small and moderate shifts and has the
form
\begin{equation}
A_{ik}\ {\mathop{=}}\left\{\hspace{0cm}
\begin{array}{ll}
{2\sum_{j=1}^{h_i}w_{ijk}\ln(w_{ijk}/(N\pi_{ij}^{(0)})) }, & {\mathrm{\ }} \quad
i=1,...,q\\[0.3cm]
(\bm {\alpha}_{i}^T{\bf
w}_{ik})^2/(N\bm{\alpha}_{i}^T\bm{\Lambda}_i\bm{\alpha}_{i}),
& {\mathrm{\ }} \quad i=q+1,...,p.
\end{array}\right.\label{f05}
\end{equation}

In fact, the above EWMA-type statistics are still heterogenous, in that they have
different distributions. To be specific, based on the property of LRTs and according
to Li et al. (2014a), in the IC state, $\frac{2-\lambda}{\lambda}A_{ik}$
asymptotically follows the chi-square distribution with $\mbox{df}(i)=h_i-1$ degrees
of freedom for nominal CDSs with $i=1,\ldots,q$ and $\mbox{df}(i)=1$ degree of
freedom for ordinal CDSs with $i=q+1,\ldots,p$. Here we employ a normalization
procedure to transform $A_{ik}$ in equation (\ref{f05}) into homogenous statistics.
Specifically, let $\chi^2_{\mbox{df}(i)}(\cdot)$ be the CDF of the chi-square
distribution with $\mbox{df}(i)$ degrees of freedom, then
\begin{align}
U_{ik}=\chi^2_{\mbox{df}(i)}\big(\frac{2-\lambda}{\lambda}A_{ik}\big),\quad
i=1,\ldots,p \label{f06}
\end{align}
will be statistics all approximately subject to the uniform distribution
$\mbox{U}(0,1)$ in the IC state. This eventually eliminates the heterogeneity
brought by various types of data streams.

Notice that the CDSs considered are assumed to be independent. As a result, $U_{ik}$
($i=1,\ldots,p$) in equation (\ref{f06}) are all independent of each other and
follow approximately the identical distribution $\mbox{U}(0,1)$ in the IC state. In
other words, they can be regarded as $p$ realizations by sampling from the uniform
distribution $\mbox{U}(0,1)$. If there are shifts in any data streams, the
calculated statistics $U_{ik}$ ($i=1,\ldots,p$) would not fit the uniform
distribution $\mbox{U}(0,1)$. This inspires us to employ a GOF test as a global test
statistic for monitoring all the $p$ CDSs simultaneously. To be specific, in the OC
state, the null hypothesis that the $p$ observations $U_{ik}$ ($i=1,\ldots,p$) are
drawn from the uniform distribution $\mbox{U}(0,1)$ should be rejected by the GOF
test. This suffices to test the hypothesis (\ref{f04}). In this sense, the
high-dimensionality is a blessing instead of a curse.

There are already many choices of GOF tests. Here we turn to the powerful GOF test
introduced by Zhang (2002), which is more powerful than many traditional GOF tests,
such as the Kolmogorov-Smirnov test and the Anderson-Darling test. This powerful GOF
test was first employed by Zou et al. (2015) for monitoring high-dimensional
continuous data streams. To be specific, let $U_{(1)k}\leq\dots\leq U_{(p)k}$ be the
ordered statistics of $U_{1k}, \dots, U_{pk}$. At time point $k$, the resulting
global monitoring statistic is
\begin{equation}
     T_k=\sum_{i=1}^p\Big[\ln\big( \frac{U_{(i)k}^{-1}-1}{(p-1/2)/(i-3/4)-1}\big)\Big]^2
     \times {\rm I}\{U_{(i)k}\geq (i-\frac{3}{4})/p\},\label{f07}
\end{equation}
where ${\rm I}\{\cdot\}$ is the indicator function. A large value of $T_k$ rejects the
null hypothesis, and hence our proposed chart would trigger an OC signal
if
\begin{equation*}
  T_k>L \ \ \mbox{for}\ \ k\geq 1,
\end{equation*}
where $L>0$ is a control limit chosen to achieve a specific  IC average
run length (ARL), denoted by $\mbox{ARL}_0$.

\section{Performance Assessment}

In this section, simulation results are presented to investigate the performance of
our proposed global monitoring statistic $T_k$ in equation (\ref{f07}). We also
compare it with two other existing approaches introduced by Tartakovsky et al.
(2006) and Mei (2010), respectively. The two methods are based on the sum and the
maximum of the local test statistics in equation (\ref{f06}) and can be expressed as
\begin{equation*}
    Q_k=\max_{i}U_{ik}\quad\mathrm{and}\quad
    S_k=\sum_{i=1}^{p}U_{ik}.
\end{equation*}

Here all the results are obtained from $10,000$ replications with the sample size
$N=100$ and the EWMA smoothing parameter $\lambda=0.1$. In addition, the $\mbox{ARL}_0$
 is set as $370$, and the control limits of the considered methods
are selected via bisection. The ARL is actually the average number of samples
required to trigger an OC signal. The OC ARLs are used for comparison. With the same
IC ARL, a smaller OC ARL indicates better performance. For the $i$th nominal CDS,
the shift in its probability vector is defined as
$\bm{\xi}_i=[\xi_{i1},\dots,\xi_{ih_i}]^T$ with $\sum_{j=1}^{h_i}\xi_{ij}=0$. For
the $i$th ordinal CDS, its latent location shift is denoted by $\delta_i$.

\begin{table}[!ht]
 \vspace{0.2cm} \center \caption{OC ARL comparison when
shifts are in one nominal case} \vspace{0.4cm}
\renewcommand{\arraystretch}{1.0}
\renewcommand\tabcolsep{4.0pt}
\begin{tabular}{r|cc|cc|cc|r|cc|cc|cc}\hline
 \multicolumn{7}{c|} {$\bm \xi_{\mathrm{a}}=[0.02,-0.02]^T$}&\multicolumn{7}{c} {$\bm \xi_{\mathrm{a}}=[0.03,-0.03]^T$}
\\\hline
   $a$ & \multicolumn{2}{c|} {$T_k$} & \multicolumn{2}{c|} {$Q_k$} & \multicolumn{2}{c|}{$S_k$}
   & $a$ &\multicolumn{2}{c|}{$T_k$} & \multicolumn{2}{c|}{$Q_k$} & \multicolumn{2}{c}{$S_k$}

\\\hline

1 & 291&(2.84)&300&(2.86) &329&(3.35) &  1 & 136&(1.21)&120&(1.07) &317&(3.06) \\
5 & 132&(1.19)&158&(1.44) &241&(2.48) &  5 & 41.4&(0.27)&46.2&(0.33) &212&(2.10) \\
10 & 74.4&(0.60)&102&(1.87) &173&(1.57) &  10 & 24.4&(0.11)&30.9&(0.17) &129&(1.17)\\
100&10.8&(0.03)&27.5&(0.14) &13.9&(0.05) &  100 & 6.44&(0.01)&13.1&(0.04) &8.35&(0.03)\\
400&4.86&(0.01)&15.6&(0.06) &5.17&(0.01) & 400 & 3.24&(0.01)&9.06&(0.03) &3.44&(0.01)\\
\hline
 \multicolumn{7}{c|} {$\bm\xi_{\mathrm{b}}=[0.02,-0.02,0]^T$}&\multicolumn{7}{c} {$\bm\xi_{\mathrm{b}}=[0.015,-0.030,0.015]^T$}
\\\hline
   $b$ & \multicolumn{2}{c|} {$T_k$} & \multicolumn{2}{c|} {$Q_k$} & \multicolumn{2}{c|}{$S_k$} & $b$ &
    \multicolumn{2}{c|}{$T_k$}
  & \multicolumn{2}{c|}{$Q_k$} & \multicolumn{2}{c}{$S_k$}
\\\hline

1 & 254&(1.53)&266&(2.57) &328&(3.30) & 1 & 162&(1.50)&151&(1.43) &329&(3.01) \\
5 & 111&(0.95)&125&(1.11) &245&(2.32) & 5 & 52.3&(0.36)&57.2&(0.44) &215&(2.04) \\
10 & 62.4&(0.46)&82.1&(0.67) &160&(1.43) & 10 &29.6&(0.15)&38.4&(0.24) &134&(1.26)\\
100&10.2&(0.03)&23.4&(0.11) &12.7&(0.05) & 100 & 7.52&(0.02)&15.3&(0.05) &9.24&(0.03)\\
300&5.48&(0.01)&15.7&(0.06) &5.69&(0.01) & 300 & 4.25&(0.01)&11.2&(0.03) &4.39&(0.01)\\

\hline
 \multicolumn{7}{c|} {$\bm \xi_{\mathrm{c}}=[0.02,0,0,-0.02]^T$}&\multicolumn{7}{c} {$\bm \xi_{\mathrm{c}}=[0.01,0.01,-0.01,-0.01]^T$}
\\\hline
   $c$ & \multicolumn{2}{c|} {$T_k$} & \multicolumn{2}{c|} {$Q_k$} & \multicolumn{2}{c|}{$S_k$} & $c$ &
    \multicolumn{2}{c|}{$T_k$}
  & \multicolumn{2}{c|}{$Q_k$} & \multicolumn{2}{c}{$S_k$}
\\\hline

1 & 229&(2.18)&232&(2.27)    &327&(3.20)     & 1 & 300&(2.83)&292&(2.79) &347&(3.44) \\
5 & 92.0&(0.78)&106&(0.94)   &232&(2.17)     & 5 & 154&(1.42)&188&(1.82) &251&(2.50) \\
10 & 51.8&(0.36)&65.7&(0.51) &156&(1.43)     & 10 &92.9&(0.75)&124&(1.11) &174&(1.63)\\
100&9.77&(0.02)&20.9&(0.09)  &11.9&(0.04)    & 100 & 12.7&(0.04)&33.2&(0.18) &14.9&(0.06)\\
300&5.30&(0.01)&14.5&(0.05)  &5.42&(0.01)    & 300 & 6.55&(0.01)&20.8&(0.09) &6.61&(0.01)\\
\hline \multicolumn{14}{l}{Note: Standard errors are in parentheses}
\end{tabular}
\end{table}

\begin{table}[!htbp]
\tabcolsep 2.5pt \vspace{-0.1cm} \centering \caption{ OC ARL comparison when shifts
are in three nominal cases} \vspace{0.4cm}
\renewcommand{\arraystretch}{1.25}
\begin{tabular}{rrr|cc|cc|cc|rrr|cc|cc|cc}\hline

   $a$&$b$&$c$&\multicolumn{2}{c|} {$T_k$} & \multicolumn{2}{c|} {$Q_k$} & \multicolumn{2}{c|}{$S_k$}
   & $a$&$b$&$c$&\multicolumn{2}{c|} {$T_k$} & \multicolumn{2}{c|} {$Q_k$} & \multicolumn{2}{c}{$S_k$}

\\\hline

            2 & 2 &  1 & 82.7&(0.68)& 86.4&(0.74)&230&(2.28)  & 40 & 40 & 20 & 9.12&(0.02)&18.9&(0.07) &11.8&(0.04)  \\
            2 & 1 &  2 & 106&(0.93)& 114&(1.01)&237&(2.19)  & 40& 20 & 40 & 10.2&(0.03)&21.9&(0.09) &13.0&(0.05)  \\
            1 & 2 &  2 & 82.0&(0.67)& 87.2&(0.72)&231&(2.29)  & 20 & 40 & 40 & 9.36&(0.02)&19.0&(0.07) &11.8&(0.04)  \\
            4 & 4 &  2 & 45.4&(0.30)& 56.2&(0.41)&162&(1.52)  & 120 & 120 & 100 & 5.03&(0.01)&13.4&(0.04) &5.31&(0.01)  \\
            4 & 2 &  4 & 58.5&(0.42)& 72.0&(0.58)&155&(1.44)  & 120 & 100 & 120 & 5.51&(0.01)&15.1&(0.05) &5.79&(0.01)  \\
            2 & 4 & 4 & 46.5&(0.31)& 56.6&(0.42)&160&(1.50)  & 100 & 120 & 120 & 5.14&(0.01)&13.5&(0.04) &5.34&(0.01)  \\
            20 & 20&10 & 13.8&(0.04)& 24.7&(0.12)&23.9&(0.13) & 200 & 200 & 100 & 3.90&(0.01)&11.5&(0.04) &3.97&(0.01)  \\
            20 & 10&20 & 15.8&(0.05)& 29.0&(0.15)&27.2&(0.16)  & 200 & 100 & 200 & 4.23&(0.01)&12.9&(0.04) &4.31&(0.01)  \\
            10 & 20&20 & 14.0&(0.04)& 24.9&(0.12)&24.4&(0.14)  & 100 & 200 & 200 & 3.98&(0.01)&11.7&(0.04) &4.36&(0.01)  \\

\hline
\multicolumn{18}{l}{Note: Standard errors are in parentheses}
\end{tabular}
\end{table}


For simplicity, first we assume that there are $p=1,000$ nominal CDSs and no ordinal
one, including three cases: (a) $400$ streams all with two attribute levels and
identical IC probabilities $[0.5,0.5]^T$; (b) $300$ ones all with three levels and IC
probabilities $[0.3,0.4,0.3]^T$; (c) $300$ ones all with four levels and IC
probabilities $[0.2,0.3,0.1,0.4]^T$. Also let $a,b,c$ be the number of changed CDSs in
each case, and further assume that in each case the shifts occurring in each
deviating stream are identical, denoted by $\bm \xi_{\mathrm{a}}$, $\bm
\xi_{\mathrm{b}}$, and $\bm \xi_{\mathrm{c}}$, respectively. Table 1 lists the
comparison results when there are shifts in only one case. From Table 1, we can see
that our proposed statistic $T_k$ is almost uniformly superior to the statistics
$S_k$ and $Q_k$, in that it possesses smaller OC ARLs in most cases. In addition,
$Q_k$ is more effective when shifts occur in only a single CDS or a few CDSs. On the
other hand, $S_k$ is more powerful when shifts occur in a large number of CDSs. The
performance of $Q_k$ and $S_k$ is expected. The former highlights the maximum among
the CDSs, which is significant if the number of shifted CDSs is small, whereas the
latter takes the sum over all the CDSs, which stands out if the number of deviating
CDSs is large. Fortunately, founded on the powerful GOF test in Zhang (2002), our
proposed $T_k$ has the advantages of both $Q_k$ and $S_k$.

In Table 2, we investigate the scenario that shifts occur simultaneously in all the
three cases. Here the shifts are set as $\bm \xi_{\mathrm{a}}=[0.020,-0.020]^T$,
$\bm \xi_{\mathrm{b}}=[0.015,-0.030,0.015]^T$, and $\bm
\xi_{\mathrm{c}}=[0.01,0.01,-0.01,-0.01]^T$. This further confirms the powerful
performance of $T_k$. Here we fix the total number of shifted CDSs. As the sum
$a+b+c$ increases, $T_k$ always stands out, $Q_k$ behaves better for small $a+b+c$,
and $S_k$ shows its advantage for large $a+b+c$.

\begin{table}[!ht]
 \vspace{0.2cm} \center \caption{OC ARL comparison when
shifts are in ordinal case} \vspace{0.4cm}
\renewcommand{\arraystretch}{1.0}
\renewcommand\tabcolsep{4.0pt}
\begin{tabular}{r|cc|cc|cc|r|cc|cc|cc}\hline
 \multicolumn{7}{c|} {$\delta_{\mathrm{d}}=0.05$} &\multicolumn{7}{c} {$\delta_{\mathrm{d}}=-0.05$}
\\\hline
   $d$ & \multicolumn{2}{c|} {$T_k$} & \multicolumn{2}{c|} {$Q_k$} & \multicolumn{2}{c|}{$S_k$} & $d$ &
    \multicolumn{2}{c|}{$T_k$}
  & \multicolumn{2}{c|}{$Q_k$} & \multicolumn{2}{c}{$S_k$}

\\\hline

               1 & 239&(2.31)& 225&(2.20)&317&(2.95)  & 1 & 240&(2.37)&225&(2.18) &318&(2.93)  \\
               5 & 95.2&(0.82)&100&(0.91) &221&(2.02)  & 5 & 94.6&(0.80)&101&(1.89) &221&(2.01) \\
               10 & 50.9&(0.36)&64.7&(0.50) &142&(1.27)  & 10 & 51.1&(0.36)&64.3&(0.51) &143&(1.28)\\
              100&9.05&(0.04)&20.4&(0.10) &11.0&(0.08) & 100 & 8.96&(0.04)& 20.3&(0.10) &11.0&(0.05)\\
               500&3.74&(0.01)&11.8&(0.04) &3.80&(0.01)  & 500 & 3.72&(0.01)&11.6&(0.04) &3.79&(0.01)\\
\hline
 \multicolumn{7}{c|} {$\delta_{\mathrm{d}}=0.10$} &\multicolumn{7}{c} {$\delta_{\mathrm{d}}=-0.10$}
\\\hline
   $d$ & \multicolumn{2}{c|} {$T_k$} & \multicolumn{2}{c|} {$Q_k$} & \multicolumn{2}{c|}{$S_k$} & $d$ &
    \multicolumn{2}{c|}{$T_k$}
  & \multicolumn{2}{c|}{$Q_k$} & \multicolumn{2}{c}{$S_k$}

\\\hline
               1 & 38.8&(0.25)& 31.8&(0.20)&310&(2.84)  & 1 & 38.0&(0.25)&32.2&(0.20) &310&(2.87)  \\
               5 & 15.2&(0.05)&16.0&(0.06) &184&(1.68)  & 5 & 15.1&(0.05)&15.9&(0.06) &184&(1.70) \\
               10 & 11.1&(0.03)&12.8&(0.04) &104&(0.97)  & 10 & 11.1&(0.03)&12.7&(0.04) &105&(0.96)\\
              100&4.06&(0.01)&7.32&(0.02) &5.10&(0.02)  & 100 & 4.05&(0.01)& 7.30&(0.02) &5.10&(0.02)\\
               500&2.00&(0.01)&5.21&(0.01) &2.01&(0.01)  & 500 & 1.99&(0.01)&5.18&(0.01) &2.01&(0.01)\\

\hline \multicolumn{14}{l}{Note: Standard errors are in parentheses}
\end{tabular}
\end{table}

\begin{table}[!htbp]
\tabcolsep 2.5pt \vspace{-0.1cm} \centering \caption{OC ARL comparison when shifts
are in one nominal or ordinal case} \vspace{0.4cm}
\renewcommand{\arraystretch}{1.0}
\begin{tabular}{r|cc|cc|cc|r|cc|cc|cc}\hline
 \multicolumn{7}{c|} {$\bm \xi_{\mathrm{a}}=[0.02,-0.02]^T$}&\multicolumn{7}{c} {$\bm \xi_{\mathrm{a}}=[0.03,-0.03]^T$}
\\\hline
   $a$ & \multicolumn{2}{c|} {$T_k$} & \multicolumn{2}{c|} {$Q_k$} & \multicolumn{2}{c|}{$S_k$}
   & $a$ &\multicolumn{2}{c|}{$T_k$} & \multicolumn{2}{c|}{$Q_k$} & \multicolumn{2}{c}{$S_k$}

\\\hline

               1 & 295&(2.90)&270&(2.62) &365&(3.44)   & 1 & 125&(1.08)&113&(0.95) &321&(2.87) \\
               5 & 132&(1.25)&156&(1.48) &233&(2.02)  & 5 & 41.4&(0.28)&45.6&(0.32) &198&(1.89) \\
               10 & 78.7&(0.64)&98.8&(0.84) &179&(1.63)  & 10 & 24.3&(0.12)&30.4&(0.17) &128&(1.21)\\
               100&10.7&(0.03)&26.7&(0.14) &13.7&(0.05)  & 100 & 6.44&(0.01)&12.9&(0.04) &8.33&(0.03)\\
               250&6.22&(0.01)&18.2&(0.08) &6.83&(0.02)  & 250 & 4.03&(0.01)&10.1&(0.03) &4.43&(0.01)\\

\hline
 \multicolumn{7}{c|} {$\bm\xi_{\mathrm{b}}=[0.01,-0.02,0.01]^T$}&\multicolumn{7}{c} {$\bm\xi_{\mathrm{b}}=[0,-0.02,0.02]^T$}
\\\hline
   $b$ & \multicolumn{2}{c|} {$T_k$} & \multicolumn{2}{c|} {$Q_k$} & \multicolumn{2}{c|}{$S_k$} & $b$ &
    \multicolumn{2}{c|}{$T_k$}
  & \multicolumn{2}{c|}{$Q_k$} & \multicolumn{2}{c}{$S_k$}
\\\hline

               1 & 320&(3.18)&285&(2.84) &322&(2.83)   & 1 & 283&(2.43)&228&(2.19) &337&(3.02) \\
               5 & 166&(1.51)&178&(1.66) &264&(2.66)  & 5 & 115&(1.03)&115&(1.06) &229&(2.14) \\
               10 & 108&(1.06)&124&(1.09) &171&(1.71)  & 10 &62.9&(0.48)&77.9&(0.63) &149&(1.29)\\
               100&13.1&(0.04)&34.4&(0.19) &15.5&(0.06)  & 100 & 10.2&(0.03)&23.0&(0.11) &12.5&(0.04)\\
               250&7.33&(0.02)&22.2&(0.10) &7.66&(0.02)  & 250 & 6.00&(0.01)&16.4&(0.04) &6.32&(0.01)\\

\hline
 \multicolumn{7}{c|} {$\bm \xi_{\mathrm{c}}=[0.01,0.01,-0.01,-0.01]^T$}&\multicolumn{7}{c} {$\bm \xi_{\mathrm{c}}=[0.01,0.01,0.01,-0.03]^T$}
\\\hline
   $c$ & \multicolumn{2}{c|} {$T_k$} & \multicolumn{2}{c|} {$Q_k$} & \multicolumn{2}{c|}{$S_k$} & $c$ &
    \multicolumn{2}{c|}{$T_k$}
  & \multicolumn{2}{c|}{$Q_k$} & \multicolumn{2}{c}{$S_k$}
\\\hline

               1 & 270&(2.24)&271&(2.45) &319&(2.87)  & 1 & 181&(1.65)&156&(1.47) & 349&(3.49)\\
               5 & 157&(1.53)&155&(1.60) &262&(2.47)  & 5 & 56.9&(0.42)&56.6&(0.41) & 210&(1.95)\\
               10 & 95.5&(0.80)&127&(1.16) &167&(1.69)  & 10 &32.2&(0.17)&38.0&(1.23) &134&(1.25)\\
               100&12.7&(0.04)&31.4&(0.17) &14.9&(0.06)  & 100 & 7.91&(0.02)&15.3&(0.05) &9.57&(0.03)\\
               250&7.31&(0.02)&21.6&(0.09) &7.37&(0.02)  & 250 & 4.87&(0.01)&12.0&(0.04) &5.03&(0.01)\\
\hline
 \multicolumn{7}{c|} {$\delta_{\mathrm{d}}=0.02$}&\multicolumn{7}{c} {$\delta_{\mathrm{d}}=-0.02$}
\\\hline
   $d$ & \multicolumn{2}{c|} {$T_k$} & \multicolumn{2}{c|} {$Q_k$} & \multicolumn{2}{c|}{$S_k$} & $d$ &
    \multicolumn{2}{c|}{$T_k$}
  & \multicolumn{2}{c|}{$Q_k$} & \multicolumn{2}{c}{$S_k$}
\\\hline
               1 & 36.6&(0.24)& 31.7&(0.20)&313&(2.89)  & 1 & 35.6&(0.23)&31.2&(0.20) &314&(2.89)  \\
               5 & 14.9&(0.05)&15.7&(0.06) &181&(1.75)  & 5 & 14.9&(0.05)&15.8&(0.06) &180&(1.75) \\
               10 & 10.9&(0.03)&12.8&(0.04) &108&(1.10)  & 10 & 10.8&(0.03)&12.7&(0.04) &108&(1.10)\\
              100&4.02&(0.01)&7.23&(0.02) &5.22&(0.01)  & 100 &4.01&(0.01)& 7.22&(0.02) &5.26&(0.01)\\
               250&2.72&(0.01)&5.94&(0.02) &2.92&(0.01)  & 250 & 2.72&(0.01)&5.94&(0.02) &2.91&(0.01)\\
\hline \multicolumn{14}{l}{Note: Standard errors are in parentheses}
\end{tabular}
\end{table}
In Table 3, we assume that there are $p=1,000$ ordinal CDSs all with four attribute
levels. Their corresponding latent continuous variables all follow the standard
normal distribution with mean 0 and variance 1, and the ordinal attribute levels are
obtained by classifying the latent variables into the four intervals
$(-\infty,-1.0],(-1.0,0.2],(0.2,0.8]$, and $(0.8,\infty)$, respectively. The shifts
in each corresponding latent continuous variable are assumed to be identical,
denoted by $\delta_{\mathrm{d}}$. Let $d$ be the number of shifted ordinal CDSs.
Throughout this table, either for positive latent location shifts or negative ones,
the proposed $T_k$ exhibits the best performance in almost all cases, and similar
patterns to Tables 1 and 2 can be observed. Please be reminded that here we limit
our attention to normally distributed latent variables only, as indicated at the end
of Section 2. If non-normal latent continuous variables of CDSs exist, we may also
transform $\alpha_{ij}$ in equation (\ref{f03a}) based on the PDF and CDF of the
standard logistic distribution, which has fair robustness against various types of
distributions.

To investigate the performance of our proposed statistic $T_k$ in more general and
complicated situations, now we consider that there are $p=1,000$ CDSs with both
nominal and ordinal ones, including four cases: (a) $250$ nominal CDSs all with two
attribute levels and identical IC probabilities $[0.5,0.5]^T$; (b) $250$ nominal ones
with three categorical levels and identical IC probabilities $[0.3,0.4,0.3]^T$; (c)
$250$ nominal ones with four categorical levels and identical IC probabilities
$[0.2,0.3,0.1,0.4]^T$; and (d) $250$ ordinal ones with the same settings as in Table
3. Accordingly, let $a$, $b$, $c$ and $d$ be the number of changed CDSs in each
case. The comparison results are listed in Table 4. Similar conclusions can be drawn
to Tables 1---3. It demonstrates that our proposed statistic $T_k$ is still the most
powerful in monitoring CDSs with heterogeneous parameters or nature, and that $T_k$
combines the advantages of both $Q_k$ and $S_k$ regardless of the total number of
changed CDSs.

\section{Case Study}

In this section, our proposed methodology is applied to a real dataset from a
semiconductor manufacturing process, which is under consistent surveillance by
monitoring variables collected from sensors. The dataset is publicly available in
the UC Irvine Machine Learning Repository
(http://archive.ics.uci.edu/ml/datasets/SECOM). Originally there are 591
measurements per observation. The dataset contains totally $1,567$ vector
observations, which were collected from July 2008 to October 2008 by a computerized
system. Among them, 104 observations that fail in the quality inspection are
classified as the nonconforming group, and the rest 1,463 observations belong to the
conforming group. Actually, not all of the 591 features contribute equally, and some
features that are constant regardless of conforming or nonconforming are removed
from the analysis. As a result, there are totally 461 features left, represented by
461 data streams. In the case of missing values, we replace them by the averages of
the observed values in the corresponding data streams and groups.

\begin{figure}[!ht]
\begin{center}
\includegraphics[width=17.0cm,height=10.0cm]{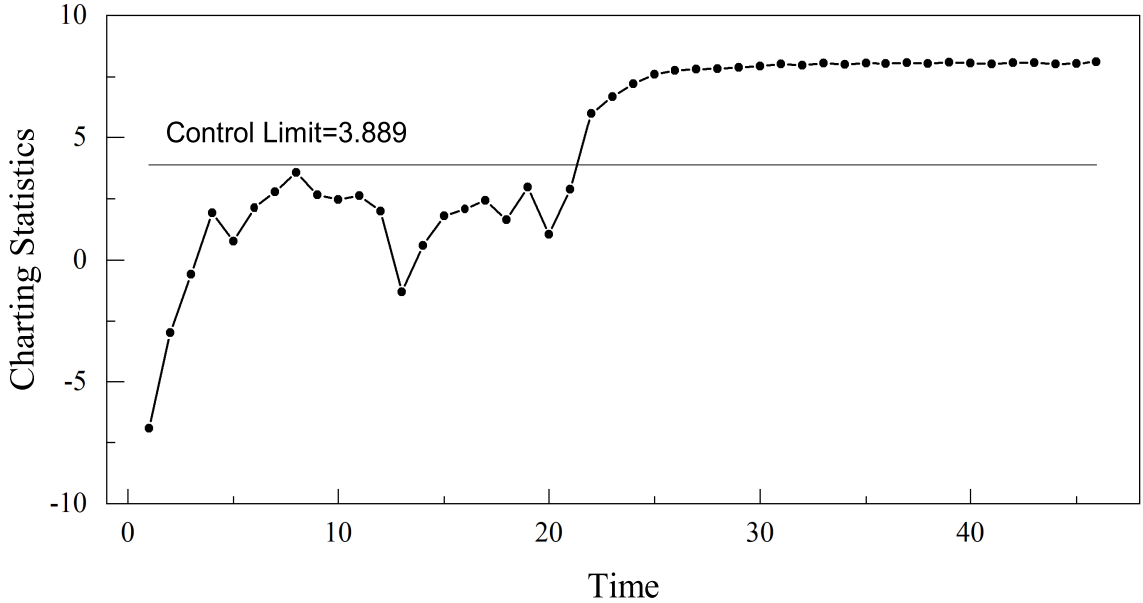}
\vspace{-0.5cm} \caption{\small $T_k$ statistics (taking logarithms) for monitoring
the semiconductor manufacturing process}
\end{center}
\end{figure}

To demonstrate the implementation of our proposed $T_k$ statistic, we first
transform the original continuous observations into categorical data. Specifically,
we regard all the 1,463 conforming vectors of dimension $461$ as the
historical IC dataset and take the averages of each variable as the thresholds that
classify the continuous values into two attribute levels. Consequentially, this
leads to 461 categorical data streams all with two levels, and we obtain the IC
probabilities of each CDS.

In Phase II, the sample size $N$ is set as four. Here we monitor the remaining 104
nonconforming observations together with 80 conforming observations. Note that they
have all been dichotomized by the thresholds calculated based on the IC dataset. In
addition, we set the IC ARL as 500 with the corresponding control limit 3.889 (after
taking the natural logarithm). The $T_k$ statistics are calculated based on equation
(\ref{f07}) and the above $20+26=46$ samples. After the logarithm operation, they
are plotted in Figure 1. According to it, this chart signals an OC alarm at the 22nd
sample and remains above the control limit.

\section{Conclusion}

In this article, we propose a powerful statistic $T_k$ to globally monitor a large
number of categorical data streams, including nominal ones with several attribute
levels and ordinal ones with some natural order among their attribute levels. Our
proposed method first eliminates the heterogeneity brought by different nature and
parameters of the data streams, and naturally integrates the local information from
each data stream based on a powerful goodness-of-fit test. Compared with two other
existing global monitoring approaches, numerical simulations reveal that the
proposed statistic is either the best or very close to the best in detecting changes
in the probability vectors of nominal CDSs or latent location shifts of ordinal
CDSs.

Although our proposed monitoring statistic is sensitive to various changes in each
CDS, diagnosing OC data streams and identifying root causes remain an open problem.
Moreover, here we assume that all CDSs shift at the same time, which is not always
the case in practice. After relaxing this assumption, how to efficiently monitor a
large number of categorical data streams simultaneously with multiple change-points
also requires future research.

\section*{References} \footnotesize \baselineskip 10pt
\begin{description}

\item {}Bersimis, S., Psarakis, S., and  Panaretos, J. (2007). Multivariate statistical
   process control charts: an overview. {\it Quality and Reliability Engineering International}, 23(5), 517--543.

\item {}Capizzi G., and Masarotto G. (2011). A Least Angle Regression Control Chart for Multidimensional Data.
{\it Technometrics}, 53(3), 285--296.

\item {}  Chiu, J., and Kuo, T. (2008). Attribute control chart for multivariate Poisson distribution.
{\it Communications in Statistics: Theory and Methods}, 37(1), 146--158.

\item {} He, S., He, Z., and Wang, G. (2014). CUSUM Control Charts for Multivariate Poisson Distribution.
{\it Communications in Statistics}, 43(6), 1192--1208.

\item {}Kamranrad, R., Amiri, A., and Niaki, S. T. A. (2017). New approaches in monitoring multivariate
categorical processes based on contingency tables in phase II.
{\it  Quality and Reliability Engineering International}, 33(5), 1105--1129.

\item {}   Li, J., Tsung, F., and Zou, C. (2012). Directional Control Schemes for Multivariate Categorical Processes.
{\it Journal of Quality Technology}, 44(2), 136--154.

\item {}   Li, J., Tsung, F., and Zou, C. (2014a). Multivariate binomial/multinomial control chart,
{\it IIE Transactions}, 46(5), 526--542.

\item {}   Li J., Tsung, F., and Zou, C.  (2014b).  A simple categorical chart for
detecting location shifts with ordinal information, {\it International Journal of Production Research}, 52(2),
550--562.

\item{} Li, J., Xu, J., and Zhou, Q. (2018). Monitoring serially dependent categorical processes with ordinal information.
{\it IISE Transaction}, 50(12), 596--605.

\item{} Liu, K., Mei, Y., and Shi, J. (2015). An adaptive sampling strategy for online high-dimensional process monitoring.
{\it Technometrics}, 57(3), 305--319.

\item{} Lu, X.S., Xie, M., Goh, T.N., and Lai, C.D. (1998). Control charts for multivariate attribute processes.
{\it International Journal of Production Research}, 36(12), 3477--3489.

\item {}Lowry, C.A., and Montgomery, D.C. (1995). A review of multivariate control charts.
{\it IIE Transactions}, 27(6), 800--810.

\item{} Marcucci, M. (1985). Monitoring multinomial processes.
{\it Journal of Quality Technology}, 17(2), 86--91.

\item{} Mei, Y. (2010). Efficient scalable schemes for monitoring a large number of data streams.
{\it Biometrika}, 97(2), 419--433.

\item{}Patel, H.I. (1973). Quality control methods for multivariate binomial and Poisson distributions.
{\it Technometrics}, 15(1), 103--112.

\item{} Tartakovsky, A. G., Rozovskii, B. L., Blazek, R. B., and Kim, H. (2006). Detection of intrusions
in information systems by sequential change-point methods. {\it Statistical Methodology}, 3(3), 252--293.

\item{} Woodall, W. H. (1997). Control charts based on attribute data: bibliography and review.
{\it Journal of Quality Technology}, 29(2), 172--183.

\item{}Wang, Z., Li, Y., and  Zhou, X. (2017). A Statistical Control Chart for
Monitoring High-dimensional Poisson Data Streams.
{\it Quality and Reliability Engineering International}, 33(2), 307--321.

\item{} Wang, K., and Jiang, W. (2009). High-Dimensional Process Monitoring and Fault Isolation via Variable Selection.
{\it Journal of Quality Technology}, 41(3), 247--258.

\item{} Zhang, J. (2002). Powerful goodness-of-fit tests based on the likelihood ratio.
{\it Journal of the Royal Statistical Society}, series B, 64(2), 281--294.

\item{} Zou, C., and Qiu, P. (2009). Multivariate Statistical Process Control Using LASSO.
{\it Publications of the American Statistical Association}, 104(488), 1586--1596.

\item{} Zou, C., Wang, Z., Zi, X., and Jiang, W. (2015). An efficient online monitoring
method for high-dimensional data streams. {\it Technometrics}, 57(3), 374--387.
\end{description}

\end{document}